# THE NEEL TEMPERATURE AND SUBLATTICE MAGNETIZATION FOR THE STACKED TRIANGULAR-LATTICE ANTIFERROMAGNET WITH A WEAK INTERLAYER COUPLING

Ignatenko A.N.[1,a], Irkhin V.Yu.[1,b] and Katanin A.A.[1,2,c]

[1] Institute of Metal Physics, Ekaterinburg, Russia

[2] Max Planck Institute for Solid State Research, Stuttgart, Germany

[a]Ignatenko@imp.uran.ru, [b]Valentin.Irkhin@imp.uran.ru, [c]A.Katanin@fkf.mpg.de

**Keywords:** two-dimensional magnetism, layered magnets, triangular lattice, renormalization group, frustration, non-linear sigma model.

**Abstract.** The quantum Heisenberg antiferromagnet on the stacked triangular lattice with the intralayer nearest-neighbor exchange interaction $J$ and interlayer exchange $J'$ is considered within the non-linear $\sigma$-model with the use of the renormalization group (RG) approach. For $J' \ll J$ the asymptotic formula for the Neel temperature $T_{\text{Neel}}$ and sublattice magnetization are obtained. RG turns out to be insufficient to describe experimental data since it does not take into account the $\mathbb{Z}_2$-vortices. Therefore $T_{\text{Neel}}$ is estimated using the Monte-Carlo result for the 2D correlation length [10] which has a Kosterlitz-type behavior near the temperature $T_{\text{KT}}$ where the vortices are activated.

## Introduction

Two-dimensional (2D) and quasi-2D spin systems have been the subject of intensive investigations during last decades. A special attention was paid to the frustrated antiferromagnets (i.e. those with competing interactions). The frustration can lead to strong non-collinear quantum and temperature-induced magnetic fluctuations. In particular, the triangular-lattice antiferromagnet (TLAF) is the most studied and interesting case since here the frustration originates simply from the geometry of the lattice. However, TLAF turns out to be the hard nut for the field-theoretical methods developed in the context of non-frustrated (i.e., collinear) antiferromagnets. In such a situation, applying these methods for real materials is instructive to test whether they lead to the correct physical picture or fail in comparison with the experiment.

In the present study we consider the application of perturbative RG to the calculation of the Neel temperature ($T_{\text{Neel}}$) and sublattice magnetization for the quasi-2D antiferromagnet with stacked triangular lattice. The Hamiltonian has the standard Heisenberg form

$$H = \sum_{\langle ij \rangle} J_{ij} \mathbf{S}_i \cdot \mathbf{S}_j, \tag{1}$$

where $\mathbf{S}_i$ are spin operators on the sites of the lattice, $\langle ij \rangle$ denotes the summation goes over nearest neighbors, $J_{ij}=J>0$ is the in-plane exchange parameter, and $0<J_{ij}=J'\ll J$ is the interlayer coupling.

## The model

It is possible to map the Hamiltonian onto the nonlinear $\sigma$-model, which is justified for low energies and long distances. This mapping was first introduced for the collinear antiferromagnetic chain by Haldane [1] and then generalized to quasi-2D case [2] and TLAF [3]. The full action of the nonlinear $\sigma$-model,

$$S_{\text{NL}\sigma} = \sum_n (S_n + S_n^{\text{int}}), \tag{2}$$

consists of the action for each layer

$$S_n = \int_0^{1/T} d\tau \int d^2x \left[ \frac{1}{2}\left(\chi_{\text{out}}^0(|\partial_\tau \mathbf{e}_1|^2 + |\partial_\tau \mathbf{e}_2|^2) - [2\chi_{\text{out}}^0 - \chi_{\text{in}}^0](\mathbf{e}_1 \partial_\tau \mathbf{e}_2)^2\right) + \right.$$
$$\left. \frac{1}{2}\left(\rho_{\text{out}}^0(|\nabla \mathbf{e}_1|^2 + |\nabla \mathbf{e}_2|^2) - [2\rho_{\text{out}}^0 - \rho_{\text{in}}^0](\mathbf{e}_1 \nabla \mathbf{e}_2)^2\right) \right], \quad (3)$$

and interaction between the layers $n$ and $n+1$

$$S_n^{\text{int}} = -\int_0^{1/T} d\tau \int d^2x \left[ \rho_{\text{out}}^0 \alpha_{\text{out}}^0 \left(\mathbf{e}_1(n)\cdot\mathbf{e}_1(n+1) + \mathbf{e}_2(n)\cdot\mathbf{e}_2(n+1)\right) + \right.$$
$$\left. \frac{1}{4}\left(\rho_{\text{in}}^0 \alpha_{\text{in}}^0 - 2\rho_{\text{out}}^0 \alpha_{\text{out}}^0\right)\left(\mathbf{e}_1(n)\cdot\mathbf{e}_2(n+1)\right)^2 \right], \quad (4)$$

where $x=(x,y)$ are spatial coordinates in a layer, $\tau$ is Matsubara-time, $\rho_{\text{out}}^0$ and $\rho_{\text{in}}^0$, $\chi_{\text{out}}^0$ and $\chi_{\text{in}}^0$, $\alpha_{\text{out}}^0 \ll 1$ and $\alpha_{\text{in}}^0 \ll 1$ are the bare spin stiffnesses, uniform susceptibilities and interlayer couplings correspondingly (subscripts "in" and "out" are related to in-plane and out-of-plane spin-wave modes in the ordered state). $\mathbf{e}_1=\mathbf{e}_1(x,\tau,n)$, $\mathbf{e}_2=\mathbf{e}_2(x,\tau,n)$ are orthonormal vectors playing role of the fluctuating order parameter, which are connected with the original spin variables in the coherent state functional integral representation by

$$\mathbf{S}_{in}(\tau) = (-1)^n S Z(\mathbf{e}_1(\mathbf{r}_i,\tau,n)\cos\mathbf{Q}\cdot\mathbf{r}_i + \mathbf{e}_2(\mathbf{r}_i,\tau,n)\sin\mathbf{Q}\cdot\mathbf{r}_i) \quad (5)$$

where $\mathbf{r}_i$ are coordinates of the lattice sites in the layer $n$, $\mathbf{Q}=(4\pi/3, 4\pi/\sqrt{3})$ is the wavevector of (long-range or short-range) magnetic order, $Z$ is some rescaling factor. As usual in the quantum-field theory, an ultraviolet momentum cutoff $\Lambda$ is introduced.

**RG equations**

The perturbative renormalization group equations for $\alpha_{\text{out}}^0$, $\alpha_{\text{in}}^0 = 0$ (i.e. for the 2D model) were derived in [4]. Here we are interested in the renormalized classical regime $T \ll \rho_{\text{out}}, \rho_{\text{in}}$ where the model becomes effectively classical with the parameters $\rho_{\text{out,in}}$, $\alpha_{\text{out,in}}$ being renormalized by quantum fluctuations and the ultraviolet cutoff $\Lambda_{\text{eff}}$ of order of $T/c_{\text{out}}$ [5]. Note that on large enough scales $\Lambda_{\text{eff}} \gg a^{-1}(J'/J)^{1/2}$ the fluctuations in the quasi-2D model are the same as in the 2D model [2]. Under these conditions we have up to two-loop order

$$\Lambda_{\text{eff}} \frac{db}{d\Lambda_{\text{eff}}} = \frac{(1+b)^2}{2y}\left[(N-1)b + 3 - N - \frac{1+b}{2y}[7(N-3) + b((N-1)b + 10 - 4N))]\right] + O(y^{-3}),$$
$$\Lambda_{\text{eff}} \frac{dy}{d\Lambda_{\text{eff}}} = (N-2)(1+b)^2\left[\frac{1}{2} + \frac{1}{8}\frac{(1+b)^2}{y}\right] + O(y^{-2}). \quad (6)$$

Here $b = \rho_{\text{in}}/\rho_{\text{out}} - 1$, $y = \rho_{\text{in}}/T$ is the inverse coupling constant, and $N$ is the number of components of the fields $\mathbf{e}_1$, $\mathbf{e}_2$ (physically $N=3$). Other RG equations for interlayer couplings and relative sublattice magnetization $\sigma = \langle\mathbf{e}_1\rangle/\langle\mathbf{e}_1\rangle_0 = \langle\mathbf{e}_2\rangle/\langle\mathbf{e}_2\rangle_0$ (symbols $\langle..\rangle$ and $\langle..\rangle_0$ denotes the statistical average with the action (2) for finite $T$ and $T=0$, respectively) can be obtained with the use of RG results [4] for the renormalization constant of the fields (see [2])

$$\Lambda_{\text{eff}} \frac{d \ln \alpha_{\text{out}}}{d \Lambda_{\text{eff}}} = \frac{3+b(2+b)}{2y} + O(y^{-2}), \quad \Lambda_{\text{eff}} \frac{d \ln \alpha_{\text{in}}}{d \Lambda_{\text{eff}}} = \frac{N-(N-2)b^2}{2y} + O(y^{-2}),$$

$$\Lambda_{\text{eff}} \frac{d \ln \sigma}{d \Lambda_{\text{eff}}} = -\frac{1}{2} \frac{(N-2)(1+b)+1}{y} + O(y^{-3}). \quad (7)$$

**The Neel temperature in the RG approach**

Eqs. 6, 7 were solved analytically. Using the matching condition $\sigma(\Lambda_{\text{eff}} \approx \sqrt{\alpha_p}) \approx 1$, $\alpha_p = \alpha_{\text{out}}(\alpha_{\text{in}}/\alpha_{\text{out}})^{1/((N-2)b+N-1)}$ which is justified at least in the first order in $1/y$, an equation for the temperature dependence of the relative sublattice magnetization $\sigma(\Lambda_{\text{eff}} \approx T/c_{\text{out}})$ was derived. Here we do not write down this equation explicitly, but present the formula for the Neel temperature,

$$T_{\text{Neel}} = 4\pi \rho_{\text{out}} \left[ K(\rho_{\text{in}}/\rho_{\text{out}}) \left( \ln \frac{2 T_{\text{Neel}}^2}{c_{\text{out}}^2 \alpha_p} + w \ln \frac{4\pi \rho_{\text{out}}}{K(\rho_{\text{in}}/\rho_{\text{out}}) T_{\text{Neel}}} \right) + C \right]^{-1}, \quad \alpha_p \to 0 \quad (8)$$

where $C$ is some unknown constant, $K(x) = N(x-x_c)^{(N-2)/(N-1)}/[2x(G(x_c)-G(x))]$, and the function $G(x)$ is given in terms of the hypergeometric function $_2F_1$

$$G(x) = x^{-N/(N-1)} {}_2F_1(N/(N-1), 1/(N-1), (2N-1)/(N-1), x_c/x) \quad (9)$$

with $x_c=2(N-2)/(N-1)$, $w=(18N^3-83N^2+112N-51)/(2(N-2)^3(2N-3))$. The Eq. 8 can be checked also by the $1/N$ expansion. In comparison with the $T_{\text{Neel}}$ result for a collinear quasi-2D antiferromagnet [4] the Eq. 8 is more complicated since we have now two independent stiffnesses. Nevertheless, it has a similar structure with the leading and sub-leading logarithmic terms in the denominator. Note that subsequent perturbative $1/y$-corrections to the right parts of the Eqs. 6,7 can modify only the constant $C$ and do not change the logarithmic terms.

**Comparison with the experiment**

In Fig. 1 the RG results for the sublattice magnetization are compared with the spin-wave theory and neutron scattering data for $VCl_2$ [6]. One can see a large disagreement between the theory and experiment. At the same time, the RG approach for the collinear antiferromagnets leads to a quantitative agreement with experiment [2]. The reason for the disagreement lies in the presence of topological $\mathbb{Z}_2$-vortices in the nonlinear $\sigma$-model Eq. 2, or, equivalently, in the Hamiltonian Eq. 1, which are obtained in the Monte-Carlo calculations [7]. It is important that the vortices are non-trivial topological configurations. These are completely neglected in perturbative RG since it catches only local properties of spin configurations, but not global ones.

The topological vortices of the Heisenberg model on the triangular lattice are to some extent similar to the vortices of the XY model [8, 9]. The main difference is that spin waves in the first model are not free, so that it is not possible to integrate them out exactly. However, the Monte-Carlo experiments [10], as well as theoretical predictions [8], show that in the 2D model the vortices at large enough distances interact with each other by the logarithmic Coulomb coupling and the correlation length $\xi(T)$ have the Kosterlitz-Thouless form [9]

$$\xi(T) = A\exp\left[b/\sqrt{T-T_{KT}}\right] \qquad (10)$$

for $T \gtrsim T_{KT}$, where $T_{KT} = 0.28\, JS^2$ and $b=0.77$ [10]. We define the Neel temperature for the quasi-2D model as a temperature where the crossover from 2D to quasi-2D regime occurs, so that

$$\xi(T_{Neel}) \approx a\sqrt{J/J'},$$
$$T_{Neel} \approx T_{KT} + 2.37 JS^2 \ln^{-2}\left(\frac{2J'}{J}\right), \qquad (11)$$

This formula is applicable for $J' \gtrsim 10^{-14} J$. For $VCl_2$ it gives $T_{Neel}=37K$ which is close to the experimental value $T_{Neel}=36K$ [6].

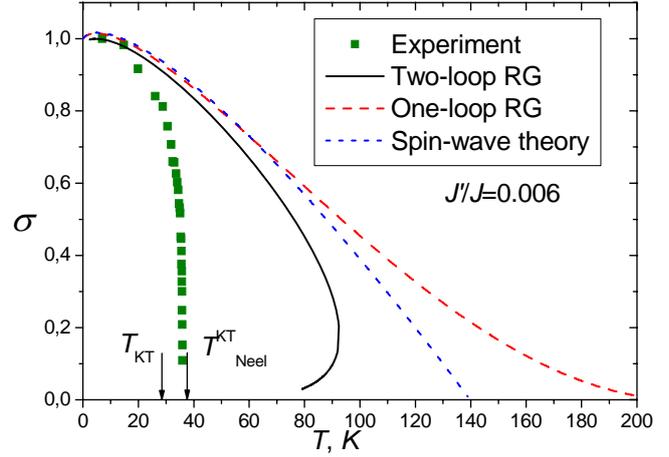

Fig. 1 The temperature dependence of relative sublattice magnetization $\sigma$. The solid, dashed and dotted lines are two-loop RG, one-loop RG and spin-wave theory results for $J'/J=0.006$, respectively. Squares are the results of the neutron scattering for $VCl_2$ [6]. $T_{KT}$ denotes the temperature where vortices are activated, and $T^{KT}_{Neel}$ is estimation of $T_{Neel}$ with account of vortices.

**Conclusion**

We conclude that perturbative RG is not sufficient to describe the Neel temperature and temperature dependence of sublattice magnetization of the real non-collinear triangular-layer antiferromagnet. This is in contrast with the situation for collinear antiferromagnet where RG works perfectly [2]. The reason for this discrepancy is the presence of the vortices which are absent for collinear antiferromagnet. The rough estimation of $T_{Neel}$ with account of vortices leads to good agreement with experiment.

The work is supported in part by the RFFI grants 07-02-01264a and 1941.2008.2 (Support of scientific schools), and the Partnergroup grant of the Max-Planck Society